\documentclass[aps,prb,twocolumn,floatfix,superscriptaddress,longbibliography]{revtex4-2}

\usepackage{graphicx}
\usepackage[utf8]{inputenc}
\usepackage{amsmath,amsfonts}

\begin{document}

\author{Ivan Abilio}
\affiliation{Department of Theoretical Solid State Physics, Institute for Solid State Physics and Optics, HUN-REN Wigner Research Center for Physics, H-1121 Budapest, Hungary} 
\affiliation{Department of Theoretical Physics, Institute of Physics, Budapest University of Technology and Economics, H-1111 Budapest, Hungary}

\author{Nicolas Néel}
\affiliation{Institut für Physik, Technische Universität Ilmenau, D-98693 Ilmenau, Germany}

\author{Jörg Kröger}
\affiliation{Institut für Physik, Technische Universität Ilmenau, D-98693 Ilmenau, Germany}

\author{Krisztián Palotás}
\email{palotas.krisztian@wigner.hun-ren.hu}
\affiliation{Department of Theoretical Solid State Physics, Institute for Solid State Physics and Optics, HUN-REN Wigner Research Center for Physics, H-1121 Budapest, Hungary} 
\affiliation{Department of Theoretical Physics, Institute of Physics, Budapest University of Technology and Economics, H-1111 Budapest, Hungary}
\affiliation{HUN-REN-SZTE Reaction Kinetics and Surface Chemistry Research Group, University of Szeged, H-6720 Szeged, Hungary}

\title{Scanning tunneling microscopy using CO-terminated probes \\ with tilted and straight geometries}

\begin{abstract}
Scanning tunneling microscopy using a CO-functionalized tip is combined with simulations to explore the impact of the CO tilt angle on topographies of a single Cu atom and CO molecule adsorbed on Cu(111)\@.
Images of the Cu atom acquired with varying tip tilt angles and sample voltages are reproduced by the calculations.
The agreement between measured and simulated data allows to unveil the tip-orbital composition of the tunneling current, which highlights the role of the different $p$-orbitals of the CO tip.
Microscope data of adsorbed CO and their dependence on voltage and probe--surface distance are captured for the nontilted junction geometry and in the limit of weak tip-surface interaction assuming sufficiently large tip--sample separations.
\end{abstract}

\maketitle

\section{Introduction}

The intentional decoration of scanning tunneling microscope (STM) or atomic force microscope (AFM) tips with a single atom or molecule is a vivid research field with substantial impact on the understanding of physics and chemistry at surfaces.
In an early experiment, chemical specificity was added to an STM W tip by its termination with an O atom \cite{prl_70_4079}.
Functionalization of the tip apex with a molecule has become a standard method to enhance the resolution of images in scanning probe experiments.
In STM experiments, their sharper electron orbitals led to submolecular resolution in images of molecules adsorbed on an insulating layer \cite{prl_80_2004,prl_94_026803,science_312_1196,science_317_1203,prl_107_086101,natchem_3_273,jpcm_29_343002,acie_57_16030,sciadv_5_eaav7717,jacs_142_13565,jacs_143_14694,science_372_852,srl_28_2140007,natchem_14_1451,acsnano_16_13092,jacs_145_967,jpcl_14_3946}.
Astounding resolution of the geometric skeleton of molecules was achieved in the Pauli repulsion distance range between tip and surface in STM \cite{njp_10_053012,prl_105_086103,jacs_133_16847} as well as in AFM \cite{science_325_1110,natchem_2_821,science_337_1326,science_340_1434,prl_113_186102,natcommun_5_3931,acsnano_10_1201,sciadv_3_e1603258,science_374_863,bjnt_10_315} experiments.
In the case of CO-terminated probes, a mechanical model revealed that the flexibility of the CO molecule and its tilting in the Pauli repulsion range is the origin of the high spatial resolution \cite{prb_90_085421}.
The propensity of a CO molecule decorating the tip to bend was later directly demonstrated by a characteristic dip-hump structure in force-versus-distance traces explored with an AFM \cite{nl_21_2318}.

The deliberate termination of STM and AFM probes is by far not restricted to the enhancement of spatial resolution.
Considerable functionality can be added as well in the spectroscopy of quantum excitations or the tracing of distance-dependent interactions.
For instance, single metal atoms at normal-metal and superconducting tips have been used to explore magnetism and magnetic excitations at the atomic limit \cite{science_350_417,sciadv_7_eabd7302}.
Very recently, metallocene molecules decorating the tip apex have been demonstrated to act as sensitive spin sensors \cite{science_364_670,science_366_623}.
In vibrational spectroscopy with an STM, the actual functionalization of the tip can profoundly influence the interpretation of the underlying inelastic electron tunneling spectroscopy data \cite{prl_87_196102,prb_76_235407,prb_83_155417,prl_110_136101,prb_93_165415,jpcl_7_2388,prl_118_036801,prl_119_166001,jpcm_31_065001}.
Even for probing the mere electronic structure of a molecular adsorbate, functionalized STM tips can be more appropriate than the ubiquitously used Tersoff-Hamann metal $s$-wave tips \cite{prl_50_1998,prb_31_805}.
Indeed, matching the tip and the adsorbate orbital symmetry led to the detection of the molecular Kondo effect, which with an $s$-wave tip was feigned to be absent in tunneling junctions \cite{prb_109_L241401}. 
Intentionally decorated tips were moreover used in AFM experiments to probe interactions between reaction partners \cite{prl_106_046104}, physisorbed and chemisorbed states \cite{science_366_235,jpca_126_6890}, relaxations in molecular contacts  \cite{njp_17_013012,njp_21_103041}, nonequilibrium bond forces \cite{nl_19_7845} as well as chemically reactive sites \cite{prl_124_096001,nature_592_722,jacs_144_7054,jpcl_13_8660}.

The interpretation of STM topographic data acquired with a CO-terminated tip is still in its infancy, despite the increasingly important role of these specific functionalized probes in contemporary imaging of surfaces at the ultimate scale.
Contributions to the tunneling current from the $s$-orbital and the three $p$-orbitals of the O atom were previously identified as important components \cite{prl_107_086101,pssb_250_2444}.
According to Chen's derivative rules \cite{prb_42_8841} obtained from the Bardeen expression of the current \cite{prl_6_57}, these contributions strongly differ.
Moreover, first-principles calculations revealed the importance of interference of electrons with different orbital origin in STM junctions \cite{prb_91_165406,prb_93_115434,prb_96_085415}.

So far, experimental and calculated STM images acquired with a CO tip relied on a special geometry at the tip apex, i.\,e., the alignment of the C-O axis with the surface normal, which will be referred to as the straight configuration in the following.
For the straight CO tip, symmetric STM and AFM images of CO molecules adsorbed on a surface were obtained \cite{prb_96_085415}.
It was speculated that decreasing the CO--CO distance would induce the bending of both molecule axes, which, however, retain their mutual parallel alignment.
Theoretical studies of the contrast origin in STM images acquired with tilted CO probes are rare due to the lack of suitable methods \cite{jpcm_25_445009,jpcm_26_485007,prb_91_165406,pss_90_223}.

The joint experiment-theory work presented here strives for filling this gap by systematically exploring measured and simulated STM images of a Cu atom and a CO molecule adsorbed on Cu(111) with CO-terminated probes that exhibit a variety of tilt angles.
In an effort to benchmark state-of-the-art theoretical approaches against experimental data, besides the variation of the tilt angle, the tunneling voltage was modified as well.
The main goal of the studies is the achievement of accordance between experimental and simulated STM images, and the explanation of specific contrast features in a qualitative and -- where possible -- in a quantitative manner.
The simulations are capable of largely reproducing the experimental data in the limit of weak tip--surface interaction and unveiling the tip-orbital composition of the tunneling current.

\section{Experimental and theoretical methods}

\subsection{Experiment}

Experiments were performed with an STM operated in ultrahigh vacuum ($10^{-9}\,\text{Pa}$) and at low temperature ($5\,\text{K}$).
The Cu(111) surface was prepared by Ar$^+$ bombardment and annealing.
Deposition of individual CO molecules was achieved by backfilling the vacuum chamber with gaseous CO (purity: $99.995\,\%$) at a partial pressure of $10^{-7}\,\text{Pa}$.
Electrochemically etched W tips were treated by field evaporation on and indentations into the Cu(111) surface, which presumably led to the coating of the tip apex with a thin Cu film.
Single Cu atoms were transferred from the tip by controlled tip--surface contacts \cite{prl_94_126102,njp_9_153,njp_11_125006,jpcm_20_223001,prl_102_086805,pccp_12_1022}.
The transfer of a single CO molecule from the surface to the tip followed a standard routine \cite{apl_71_213}.
Approaching to and retracting from clean Cu(111), these CO-terminated tips exhibit a reversible bending for voltages $\vert V\vert\leq 0.5\,\text{V}$ and currents $I\leq 500\,\text{nA}$ \cite{nl_21_2318}, which reveals their structural stability.
Constant-current and constant-height modes with the voltage applied to the sample were applied for the acquisition of STM data, which were further processed with WSXM \cite{rsi_78_013705}.

\subsection{Theory}
\label{sec:theory}

Calculations were performed in the non-spin-polarized version of density functional theory (DFT) using the Vienna Ab initio Simulation Package \cite{prb_54_11169} within the Perdew-Burke-Ernzerhof parametrization \cite{prl_77_3865} of the generalized gradient approximation with projector augmented waves \cite{prb_59_1758} and an energy cutoff of the plane wave expansion of the electron wave functions of $400\,\text{eV}$\@.
Supercell slabs are composed of five atomic layers of Cu(111) with an in-plane lattice constant of $0.257\,\text{nm}$, each layer containing $4 \times 4$ Cu atoms, on which a Cu atom or an upright CO molecule is adsorbed on one side of the Cu slab.
The adsorbed Cu atom resides at a hollow site \cite{surfsci_323_71} while the CO molecule occupies an on-top site of Cu(111) with the C-O axis aligned with the surface normal \cite{surfsci_161_349}.
A separating vacuum region of $1.67\,\text{nm}$ (Cu) and $1.55\,\text{nm}$ (CO) was used to avoid unphysical interactions between the repeated image slabs in the perpendicular Cu(111) direction.
The bottom three Cu atomic layers were fixed and all other atoms were freely relaxed during the geometry optimizations, with a convergence criterion for the forces acting on individual atoms set to $0.2\,\text{eV}/\text{nm}$ (Cu) and $0.1\,\text{eV}/\text{nm}$ (CO)\@.
The Brillouin zone was sampled by a $4 \times 4 \times 1$ $\Gamma$-centered Monkhorst–Pack k-point mesh \cite{prb_13_5188} both during geometry optimization and the subsequent calculation of the electronic states for the STM simulations.

Functionalized CO tips were modeled with the same slab model as the CO--Cu(111) surface structure described above. 
Rather than tilting the CO axis, the tilted CO tips were simulated within the revised Chen's derivative rules \cite{prb_91_165406} by rotating the electron orbitals of the O apex atom (see below)\@.

Calculations of the tunneling current and simulations of STM images were performed with the revised Chen method \cite{prb_91_165406} as implemented in the BSKAN code \cite{pss_71_147,jpcm_17_2705}. 
According to Bardeen's expression for the tunneling current, the current $I$ at bias voltage $V$ reads \cite{prl_6_57}
\begin{equation}
I\propto\sum\limits_{\mu\nu}f(E_{\mu})[1-f(E_{\nu}+eV)]\vert M_{\mu\nu}\vert^2\delta(E_{\mu}-E_{\nu}-eV),
\label{eq:current}
\end{equation}
where $f$ is the Fermi-Dirac distribution function, $\delta$ is the Dirac delta function, and the indices $\mu$ and $\nu$ denote sample and tip states with Kohn-Sham eigenenergies $E_{\mu}$ and $E_{\nu}$ corresponding to single-electron wave functions of the sample ($\psi_{\mu}$) and the tip. 
The tunneling matrix element $M_{\mu\nu}$ is calculated as a linear combination of spatial derivatives of the single-electron surface wave function and evaluated at the tip apex position $\mathbf{r}_0$ according to the orbital symmetry $\beta\in\{s,p_x,p_y,p_z\}$ of the O apex atom of the tip,
\begin{equation}
M_{\mu\nu}=\sum\limits_{\beta}M_{\mu\nu\beta}=\sum\limits_{\beta}C_{\nu\beta}\hat{\partial}_{\nu\beta}\psi_\mu(\mathbf{r}_0).
\label{eq:M}
\end{equation}
Here, $C_{\nu\beta}$ are the coefficients of the linear combination, which are generally energy-dependent complex numbers, and $\hat{\partial}_{\nu\beta}$ are differential operators corresponding to Chen's derivative rules \cite{prb_42_8841}, i.\,e., $\hat{\partial}_{\nu\beta}=1,\kappa_\nu^{-1}\partial/\partial x,\kappa_\nu^{-1}\partial/\partial y,\kappa_\nu^{-1}\partial/\partial z$ corresponding to the $s$, $p_x$, $p_y$, $p_z$-orbital of the O tip apex atom, respectively, with $\kappa_\nu$ the vacuum decay constant of the tip orbital. 
The tunneling transmission can be expressed as \cite{prb_91_165406}
\begin{equation}
\vert M_{\mu\nu}\vert^2\propto\frac{1}{\kappa_\nu^{2}}\left[\sum\limits_\beta\vert M_{\mu\nu\beta}\vert^2+\sum\limits_{\beta\neq\beta'}2\Re(M_{\mu\nu\beta}^\ast M_{\mu\nu\beta'})\right].
\label{eq:M2}
\end{equation}
Owing to this mathematical construction of the transmission, the tunneling current in Eq.\,(\ref{eq:current}) can be recast in terms of tip-orbital contributions as
\begin{equation}
I=\sum\limits_{\beta\beta'}I_{\beta\beta'}=\sum\limits_\beta I_{\beta\beta}+\sum\limits_{\beta\neq\beta'}I_{\beta\beta'},
\label{eq:sumIbeta}
\end{equation}
where $I_{\beta\beta'}=I_{\beta'\beta}$ represent symmetric current composition matrices of dimension $4\times 4$ for the mentioned choice of O orbitals.
The diagonal elements of such matrices correspond to direct contributions to the current in terms of tip orbitals (positive real values), and the off-diagonal elements reflect tip-orbital interference (positive or negative real values)\@.

To identify the most appropriate CO tip in the simulations, different tip models were used.
The DFT-derived CO tip exhibits a linear combination of O orbitals with energy-dependent $C_{\nu\beta}$ expansion coefficients.
In addition, a pure $s$-wave tip as well as $(1/\sqrt{2})(s+p_z)$-wave and $(1/2)(s+p_x+p_y+p_z)$-wave tips with equal and constant, i.\,e., energy-independent expansion coefficients, entered the simulations in order to check their validity with respect to the DFT-derived CO tip and to explore their performance in reproducing experimental data.

\begin{figure}
\centering
\includegraphics[width=0.95\columnwidth]{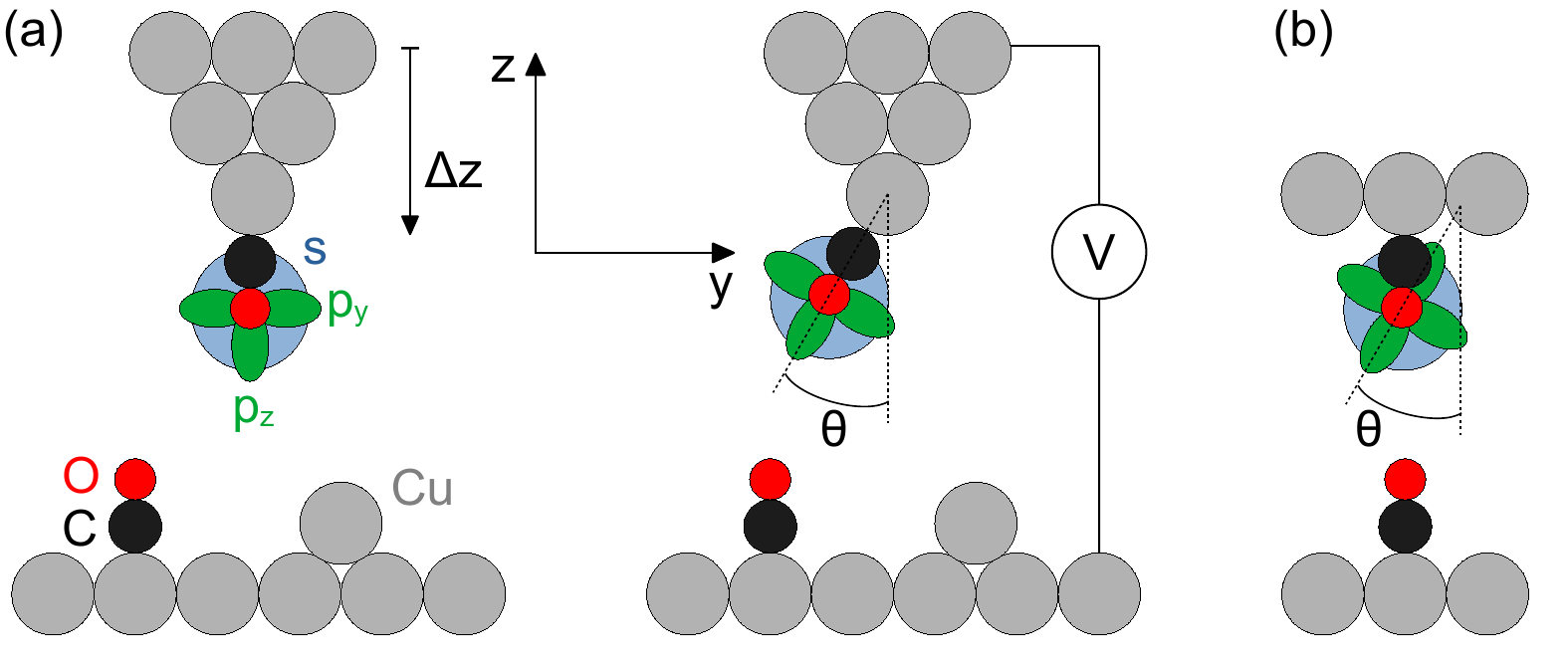}
\caption{(a) Sketch of the STM junction used in this work with marked tip displacement $\Delta z$, sample voltage $V$, and CO molecule as well as Cu atom adsorbed on Cu(111)\@.
Electron orbitals of the O atom of the CO-decorated tip are schematically illustrated as a circle ($s$) and elliptical lobes ($p_y$, $p_z$, the $p_x$-orbital sticks out of the indicated $yz$ plane and is omitted here)\@. 
The $y$ direction is oriented along a compact $\langle 11\bar{2}\rangle$ direction of the Cu(111) lattice while the $z$ axis coincides with the surface normal.
(b) Illustration of the main concept of the simulations.
The tilted tip is modeled as rigidly rotated electron orbitals of the O apex atom \cite{prb_91_165406} (see text for more details)\@. 
The tilt angle $\theta$ of CO at the tip apex is a parameter of the simulations.}
\label{fig1}
\end{figure}

Tilted tips are considered by redefined coefficients of the linear combination of spatial derivatives of $\psi_\mu$ within the revised Chen method \cite{prb_91_165406}, where the electron orbital structure of the O tip apex atom is rigidly rotated into the new tip coordinate system upon tilting \cite{jpcm_25_445009,jpcm_26_485007,pss_90_223}. 
The coordinate transformation matrix relating the coordinates of the tilted tip $x'^j\in\{x',y',z'\}$ to the sample coordinates $x^i\in\{x,y,z\}$ with tilt angle $\theta$ corresponding to a rotation with respect to the $x$ axis (Fig.\,\ref{fig1}) is
\begin{equation}
R=
\begin{bmatrix}
    1 & 0 & 0 \\
    0 & \cos\theta & \sin\theta \\
    0 & -\sin\theta & \cos\theta
\end{bmatrix},
\end{equation}
with $x'^j=R^j_i x^i$ and $x^i=\left(R^{-1}\right)^i_j x'^j$, following the Einstein summation convention. 
Consequently, the first spatial derivatives of $\psi_\mu$ corresponding to the O $p$-orbitals are \cite{prb_91_165406}
\begin{equation}
\frac{\partial\psi_\mu}{\partial x'^j}=\frac{\partial\psi_\mu}{\partial x^i}\left(R^{-1}\right)^i_j.
\end{equation}
As a result, the spatial derivatives in the tilted coordinate system of the tip are expressed in the coordinates of the sample as $\hat{\partial}'_{\nu\beta}=1,\kappa_\nu^{-1}\partial/\partial x,\kappa_\nu^{-1}(\cos\theta\,\partial/\partial y+\sin\theta\,\partial/\partial z),\kappa_\nu^{-1}(-\sin\theta\,\partial/\partial y+\cos\theta\,\partial/\partial z)$ for the tilted orbitals of the O tip apex atom $s'=s$, $p'_x=p_x$, $p'_y$, $p'_z$, respectively.

The main advantage of the described approach is the restriction to a single DFT calculation of a straight CO molecule on Cu(111) to model various tilted CO tips within the revised Chen’s derivative rules \cite{prb_91_165406}.
This rotated tip orbital approach has successfully been cross-checked against the mature approaches due to Bardeen \cite{prl_6_57} as well as Tersoff and Hamann \cite{prl_50_1998,prb_31_805}, revealing its validity \cite{jpcm_25_445009,jpcm_26_485007,prb_91_165406,pss_90_223}.
Importantly, the deployed STM model assumes the limit of weak coupling between tip and sample. Therefore, atomic relaxations in the STM junction and interactions between the CO molecule at the tip and the atomic structure of the Cu adatom or the CO molecule on Cu(111) are not taken into account.

In the STM simulations a three-dimensional current map is calculated, from which constant-current contours are extracted. 
The real tip height above the adsorbate measured from the surface atomic layer of the Cu(111) substrate can be obtained by adding the adsorbate height ($187\,\text{pm}$ for the Cu adatom, $301\,\text{pm}$ for the CO molecule) to the reported initial tip--adsorbate distances (see below)\@.
The best match of simulated with measured STM data served as the basis for the ensuing tip-orbital decomposition of the tunneling current.
The latter is represented by fractional current composition matrices, where each matrix element refers to a fractional contribution, $\tilde{I}_{\beta\beta'}=I_{\beta\beta'}/\sum\limits_{\beta\beta'}I_{\beta\beta'}$.
The sum of all matrix elements $\sum\limits_{\beta\beta'}\tilde{I}_{\beta\beta'}$ is therefore unity ($100\,\%$), and reported deviations from this are due to rounding errors.

\section{Results and discussion}

\subsection{Junction geometry}

The basic tunneling junction used in the experiments [Fig.\,\ref{fig1}(a)] and simulations [Fig.\,\ref{fig1}(b)] comprises a CO-terminated tip with a tilt angle $\theta$ subtending the surface normal in the $yz$-plane.
The left panel of Fig.\,\ref{fig1}(a) depicts the situation for a straight CO tip ($\theta=0^\circ$)\@.
Besides the tilt angle, the tip--surface distance ($z$) is changed by displacing the tip ($\Delta z$), and the voltage $V$ applied across the junction is modified [right panel of Fig.\,\ref{fig1}(a)]\@.
The $s$-orbital and two $p$-orbitals of the O atom are depicted as circles and elliptical lobes, respectively. 

\subsection{Adsorbed Cu atom}

\begin{figure}
\centering
\includegraphics[width=0.95\columnwidth]{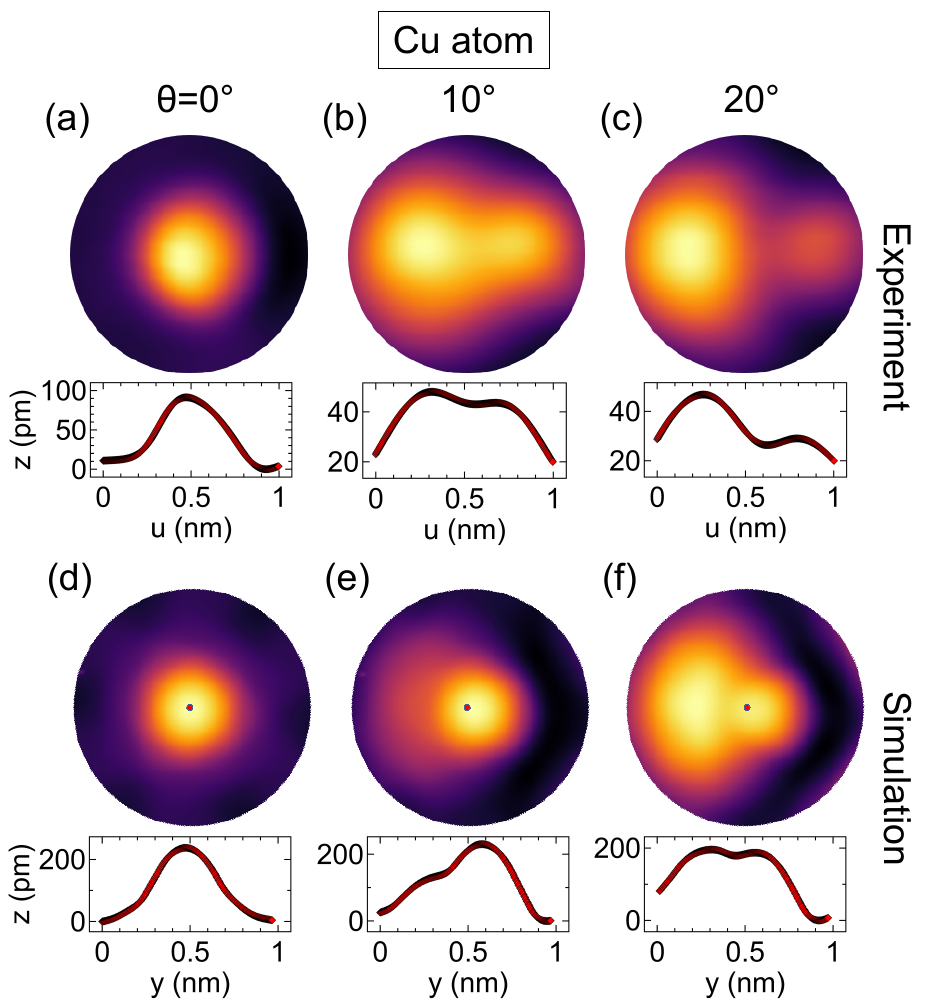}
\caption{(a)--(c) Constant-current STM images of a Cu adatom on Cu(111) recorded with a CO--terminated tip (sample voltage: $100\,\text{mV}$, tunneling current: $50\,\text{pA}$)\@.
The color scales cover apparent heights from $0$ (black) to (a) $84\,\text{pm}$, (b) $48\,\text{pm}$, (c) $47\,\text{pm}$ (yellow)\@.
(d)--(f) Simulated constant-current STM images using an ($s+p_{z}$)-wave tip (sample voltage: $100\,\text{mV}$, initial tip--adsorbate distance: $500\,\text{pm}$) with indicated tilt angles.
The color scales cover apparent heights from $0$ (black) to (d) $260\,\text{pm}$, (e) $230\,\text{pm}$, (f) $200\,\text{pm}$ (yellow)\@.
Dots mark the position of the Cu atom on the surface.
The diameter of the STM images is $1\,\text{nm}$ in (a)--(f)\@.
All cross-sectional profiles depict data along a horizontal line through the center of the STM images with $u$ ($y$) measuring the distance in the $xy$-plane (along $\langle 11\bar{2}\rangle$)\@.}
\label{fig2}
\end{figure}

The transfer of a single CO molecule adsorbed on Cu(111) to the tip results in a CO-terminated tip with the CO axis tilted by an angle $\theta$ with respect to the surface normal [Fig.\,\ref{fig1}(a)]\@.
Preparing a multitude of CO tips gives rise to a statistical distribution of tilt angles with no identifiable preference for a specific angle \cite{nl_21_2318}.
Importantly, except for the straight ($\theta=0^\circ$) adsorption of CO on the tip, tilt angles cannot be determined from the experimental data due to the unknown tip--surface distance.
Qualitatively, the tilt angle is larger the farther the protrusions are separated in STM images of the adsorbed atom (adatom) [Fig.\,\ref{fig2}(a)--(c)]\@.
Therefore, simulations of STM images were performed for $\theta=0^\circ, 10^\circ, 20^\circ$ (and beyond)\@.
Comparing then a selected experimental STM image with all available simulated data gave rise to a best match and allowed the assignment of the tilt angle of the calculations to the experimental data, as demonstrated for a specific example in the Appendix; that is, quoting the specific values of $\theta=10^{\circ}$ and $\theta=20^{\circ}$ in the following relies on the best match of measured STM images with simulations performed at these two angles.

Figure \ref{fig2} compares experimental [Fig.\,\ref{fig2}(a)--(c)] and calculated [Fig.\,\ref{fig2}(d)--(f)] constant-current STM images of a single Cu adatom on Cu(111) acquired with a CO tip at a given sample voltage ($100\,\text{mV}$) and with varying tilt angles $\theta\in\{0^\circ,10^\circ,20^\circ\}$\@.
In the experiments, the Cu adatom gives rise to a single protrusion with nearly circular circumference at $\theta=0^\circ$ [Fig.\,\ref{fig2}(a)]\@.
Its apparent height is $h\approx 84\,\text{pm}$ and full width at half maximum (FWHM) $w\approx 0.40\,\text{nm}$.
With increasing $\theta$, the STM images show two protrusions, where the larger one has an elliptical and the smaller one a circular shape.
Their brightness in STM data changes with $\theta$.
The apparent height is $\approx 48\,\text{pm}$ and $\approx 44\,\text{pm}$ ($\approx 47\,\text{pm}$ and $\approx 30\,\text{pm}$), while the FWHM (in $y$ direction, left to right in Fig.\,\ref{fig2}) is $\approx 0.70\,\text{nm}$ and $\approx 0.47\,\text{nm}$ ($\approx 0.71\,\text{nm}$ and $\approx 0.44\,\text{nm}$) for the two protrusions at $\theta=10^\circ$ ($\theta=20^\circ$)\@.
Their mutual distance increases with increasing tilt angle from $\approx 0.38\,\text{nm}$ at $\theta=10^\circ$ [Fig.\,\ref{fig2}(b)] to $\approx 0.53\,\text{nm}$ at $\theta=20^\circ$ [Fig.\,\ref{fig2}(c)]\@.

\begin{figure}
\centering
\includegraphics[width=0.95\columnwidth]{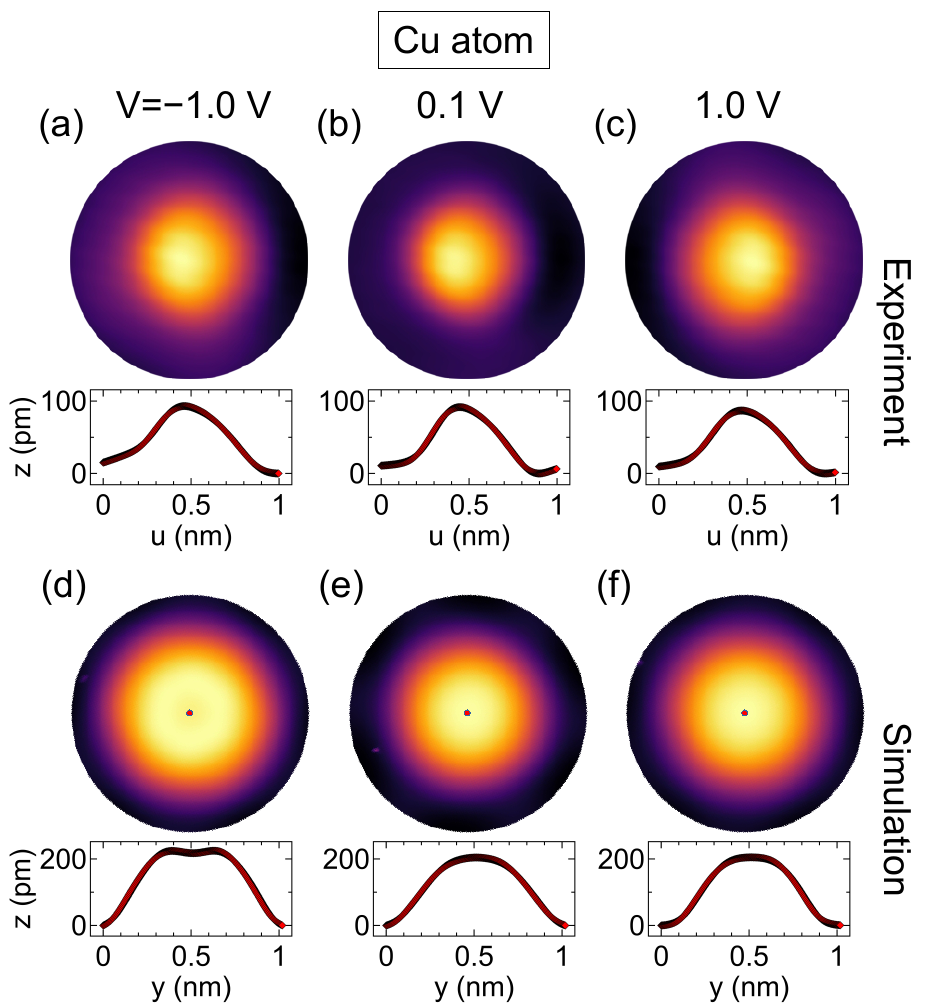}
\caption{(a)--(c) Constant-current STM images of a Cu adatom adsorbed on Cu(111) recorded with a straight CO-terminated tip at indicated sample voltages ($50\,\text{pA}$)\@.
The color scale covers apparent heights ranging from $0$ (black) to $92\,\text{pm}$ (yellow)\@.
(d)--(f) Simulated constant-current STM images using the DFT-derived straight CO tip (initial tip--adsorbate distance: $300\,\text{pm}$)\@.
The color scale covers apparent heights ranging from $0$ (black) to $235\,\text{pm}$ (yellow)\@.
Dots mark the position of the Cu adatom.
The diameter of the STM images is $1\,\text{nm}$ in (a)--(f)\@.
All cross-sectional profiles depict data along a horizontal line through the center of the STM images.}
\label{fig3}
\end{figure}

A consistent best match of simulated STM images [Fig.\,\ref{fig2}(d)--(f)] with experimental topographies has been achieved with an ($s+p_z$)-wave tip.
While the actual absolute value for apparent heights, FWHM and protrusion separations differ from the experimental data, the observed trend is reproduced (vide infra)\@.
In particular, the shape of the adjacent protrusions for $\theta=10^\circ,20^\circ$ matches well the experimental observation.
Other tips have likewise been tested in the simulations, as exemplarily shown for the data set of Fig.\,\ref{fig2}(b) in the Appendix.
The spherically symmetric O $s$-orbital expectedly led to a single protrusion in simulated STM images of the adatom for both straight and tilted tips.
The $(s+p_x+p_y+p_z)$-wave tip gave rise to a double-feature contrast of the adatom in the straight-tip configuration (Appendix), which was removed in the tilted-tip geometry, in clear conflict with the experimental observation. 
In addition, while the DFT-derived straight CO tip provided good agreement with experimental data (Fig.\,\ref{fig3}), its tilted counterpart failed to reproduce the observed double-feature contrast.

Imaging the single adatom with a straight $(s+p_z)$-wave CO tip gives rise to an apparent height of $h\approx 260\,\text{pm}$ and an FWHM of $w\approx 0.40\,\text{nm}$.
Upon tilting the tip, two protrusions occur in the simulated STM images with apparent heights of $\approx 110\,\text{pm}$ and $\approx 230\,\text{pm}$ ($\approx 200\,\text{pm}$ and $\approx 185\,\text{pm}$), and FWHM of $\approx 0.40\,\text{nm}$ and $\approx 0.31\,\text{nm}$ ($\approx 0.54\,\text{nm}$ and $\approx 0.29\,\text{nm}$) at $\theta=10^\circ$ ($\theta=20^\circ$)\@.
The splitting of the single protrusion at $\theta=0^\circ$ increases from $\approx 0.31\,\text{nm}$ ($\theta=10^\circ$) to $\approx 0.37\,\text{nm}$  ($\theta=20^\circ$)\@.
Using the orbital decomposition of the tunneling current discussed below, the left protrusion results from the overlap of the O $s$-orbital with the adatom-surface wave function, while the right protrusion can be assigned to the interaction of the tilted O $p_z$-orbital with the wave function of the adatom-surface hybrid.

\begin{figure}
\centering
\includegraphics[width=0.95\columnwidth]{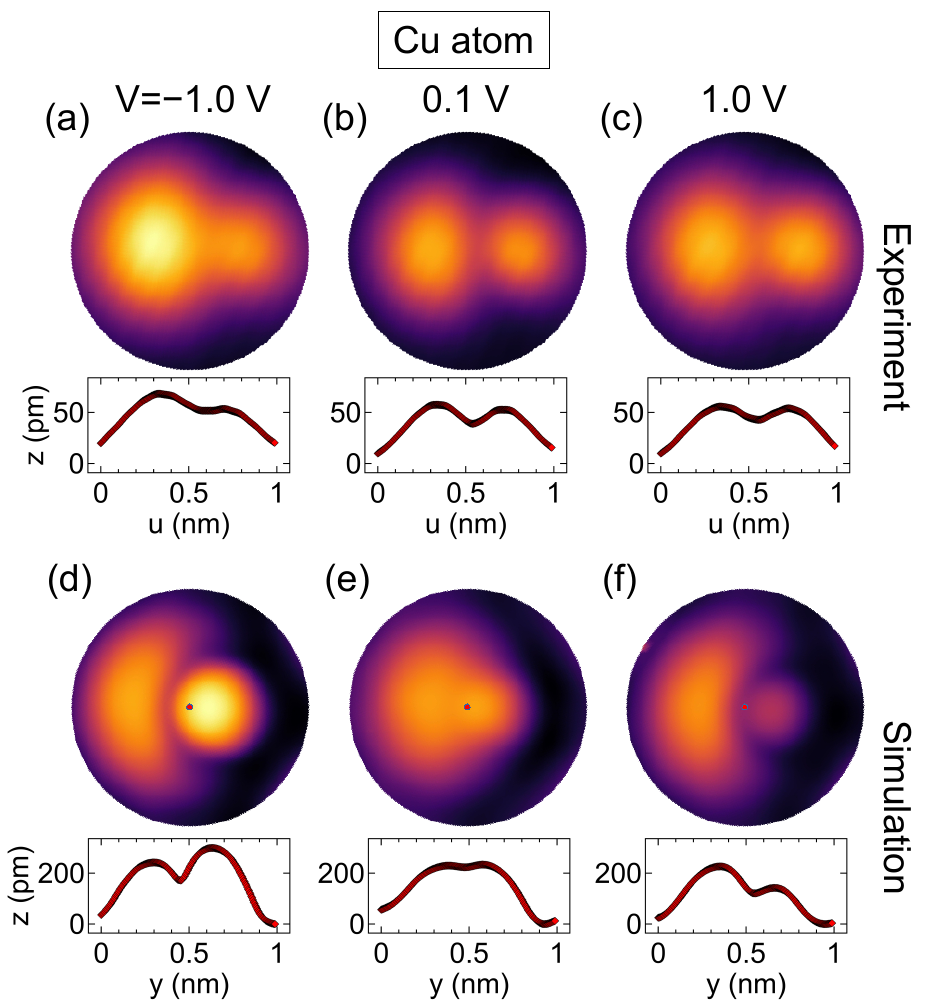}
\caption{(a)--(c) Constant-current STM images of a Cu adatom adsorbed on Cu(111) recorded with a tilted CO-terminated tip at indicated sample voltages ($50\,\text{pA}$)\@.
The color scale covers apparent heights ranging from $0$ (black) to $68\,\text{pm}$ (yellow)\@.
(d)--(f) Simulated constant-current STM images using a ($s+p_z$)-wave tip with tilt angle $\theta=20^\circ$ (initial tip--adsorbate distance: $300\,\text{pm}$)\@.
The color scale covers apparent heights ranging from $0$ (black) to $300\,\text{pm}$ (yellow)\@.
Dots mark the position of the Cu adatom.
The diameter of the STM images is $1\,\text{nm}$ in (a)--(f)\@.
All cross-sectional profiles depict data along a horizontal line through the center of the STM images.}
\label{fig4}
\end{figure}

In a next step, the voltage dependence of measured and calculated STM images was explored for $\theta=0^\circ$ (Fig.\,\ref{fig3}) and $\theta=20^\circ$ (Fig.\,\ref{fig4})\@.
The overall changes in the explored sample voltage range $\vert V\vert\leq 1\,\text{V}$ are small in both experimental and simulated data.
Cross-sectional profiles were again used to compare apparent heights and FWHM\@.
For $\theta=0^\circ$ in the experiments, $h\approx 92\,\text{pm}$ and $w\approx 0.44\,\text{nm}$ at $-1.0\,\text{V}$ were obtained [Fig.\,\ref{fig3}(a)]\@.
At $0.1\,\text{V}$ and $1.0\,\text{V}$, $h$ varies from $\approx 84\,\text{pm}$ to $\approx 87\,\text{pm}$, while $w$ essentially stays constant ($\approx 0.42\,\text{nm}$)\@. 
A Cu adatom imaged with a metal tip appears approximately half as high in the same voltage range.
The simulations with a straight CO tip give rise to $h\approx 235\,\text{pm}$, $w\approx 0.69\,\text{nm}$ ($-1.0\,\text{V}$), $h\approx 220\,\text{pm}$, $w\approx 0.57\,\text{nm}$ ($0.1\,\text{V}$), and $h\approx 210\,\text{pm}$, $w\approx 0.62\,\text{nm}$ ($1.0\,\text{V}$)\@.

For the tilted CO tip (Fig.\,\ref{fig4}) the apparent height of the broad protrusion decreases from $h=68\,\text{pm}$ at $-1.0\,\text{V}$ to $h=58\,\text{pm}$ at $0.1\,\text{V}$ and $1.0\,\text{V}$, while an apparent height of $h=53\,\text{pm}$ for the sharp protrusion is virtually independent of the voltage.
Simulated images using an ($s+p_z$)-wave tip reproduce the evolution of the broad protrusion, with an apparent height being $15\,\text{pm}$ larger at $-1.0\,\text{V}$ than at $0.1,\text{V}$ and $1.0\,\text{V}$\@.
The sharp protrusion presents a monotonic decrease of $\approx 150\,\text{pm}$ from $-1.0\,\text{V}$ to $1.0\,\text{V}$, which is at odds with the experiment. 
Overall, the ($s+p_z$)-wave tip provided the best match of the simulations with the experiments.

\begin{figure*}
\centering
\includegraphics[width=0.8\textwidth]{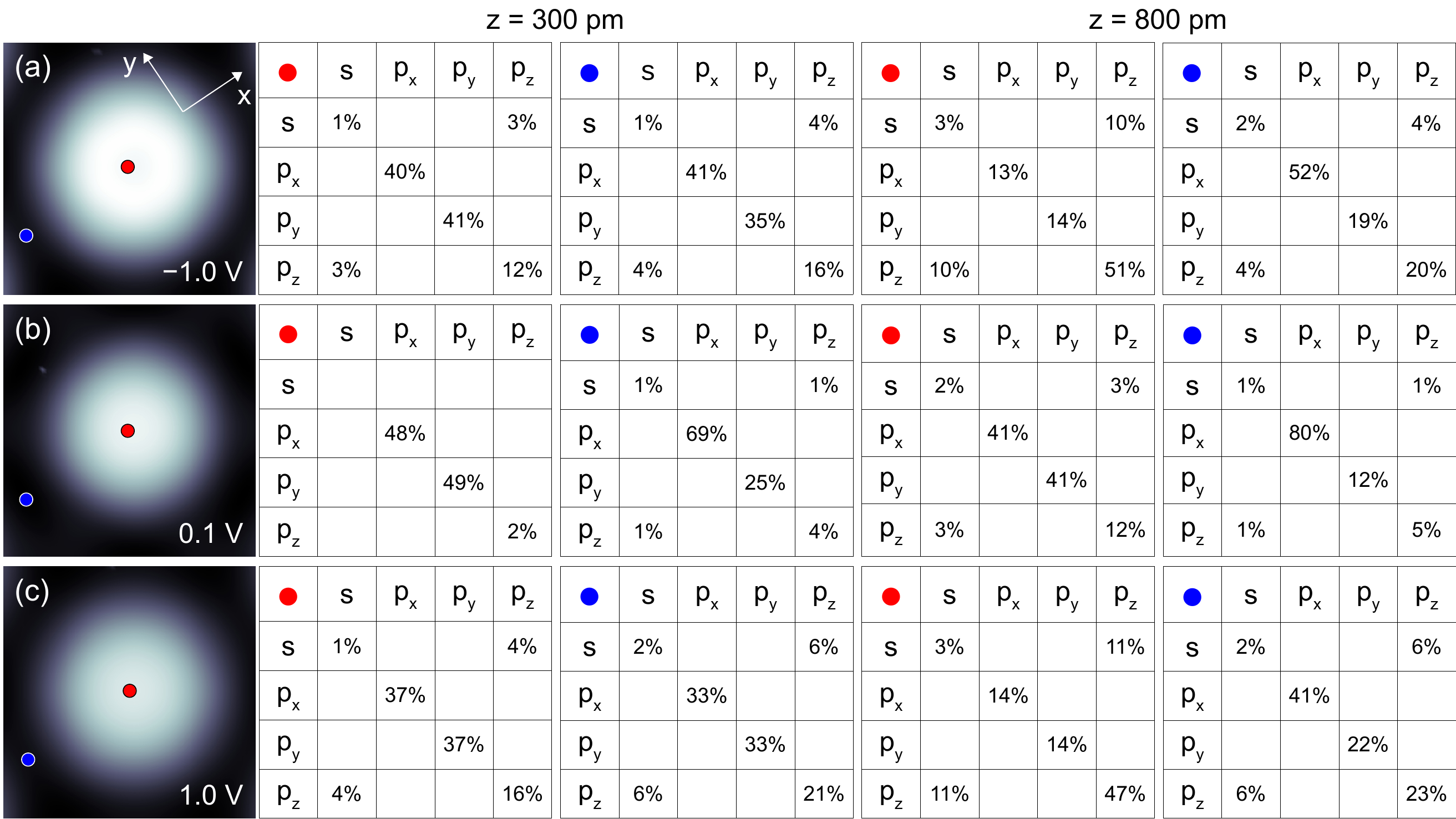}
\caption{Tip orbital decomposition of the tunneling current at sample voltages (a) $-1.0\,\text{V}$, (b) $0.1\,\text{V}$ and (c) $1.0\,\text{V}$ for a Cu adatom on Cu(111) imaged with the straight DFT-derived CO tip at $z = 300\,\text{pm}$ and $z = 800\,\text{pm}$.
Lateral tip positions are indicated by the colored dots in the simulated STM images ($1\,\text{nm}\times 1\,\text{nm}$)\@.
Values are given as fractions ($\%$) of the total current, empty entries reflect vanishing contributions.
}
\label{fig5}
\end{figure*}

The overall good agreement between experimental and calculated data for the Cu adatom imaged with a straight CO tip at different voltages (Fig.\,\ref{fig3}) encouraged the analysis of the tip-orbital composition of the tunneling current.
To this end, the revised Chen's method \cite{prb_91_165406} was used to decompose the current in terms of the real-space electron orbitals of the O apex atom of the DFT-derived CO tip following Eq.\,(\ref{eq:sumIbeta})\@.
Figure \ref{fig5} shows the results for sample voltages $-1.0\,\text{V}$ [Fig.\,\ref{fig5}(a)], $0.1\,\text{V}$ [Fig.\,\ref{fig5}(b)], $1.0\,\text{V}$ [Fig.\,\ref{fig5}(c)], and for tip--surface distances of $z=300\,\text{pm}$, $z=800\,\text{pm}$ above two different sites (center of and separated from the adatom)\@.

Above the Cu adatom center, the O $p_z$-orbital contributes significantly to the current at $\pm 1.0\,\text{V}$ and becomes even dominant for large tip--surface distances at these voltages.
The polarity of the voltage does not affect this behavior.
At $0.1\,\text{V}$, the tunneling current is instead carried by O $p_x$, $p_y$-orbitals, which is in line with their dominant density of states (DOS)  at the Fermi energy ($E_\text{F}$)\@.
These tip orbitals efficiently overlap with the $s$, $d_{xz}$, $d_{yz}$, $d_{z^2}$-orbitals of the Cu adatom, which are most important for the partial DOS of the surface at $E_\text{F}$\@.
Upon increasing the tip--surface separation, at $\pm 1.0\,\text{V}$ the contribution of $p_x$ and $p_y$ becomes less important for the benefit of the $p_z$-orbital, which now overlaps with the $s$, $d_{z^2}$-orbital of the Cu adatom.

The split protrusion appearing when imaging the Cu adatom with a tilted CO probe can be understood as follows. 
The CO tip is tilted in the $yz$-plane along the negative $y$-axis (Fig.\,\ref{fig1}) and scans along the $y$-axis from negative to positive values with the Cu adatom residing at $y=0$\@.
A first protrusion in constant-current STM images develops when $s$-orbitals of the CO molecule and the Cu adatom overlap.
The CO $p_z$-orbital plays a minor role because it is tilted away from the adatom site.
After surmounting and passing the Cu adatom, the overlap of $s$-orbitals reduces giving rise to a decreasing tunneling current and a concomitantly decreasing apparent height in the constant-current STM image.
The apparent height increases again on the verge of interaction between the CO $p_z$-orbital and the Cu atom $s$-orbital, which for the continued scan along the $y$-axis causes the second protrusion in the STM image.

\subsection{Adsorbed CO molecule}

\begin{figure}
\centering
\includegraphics[width=0.95\columnwidth]{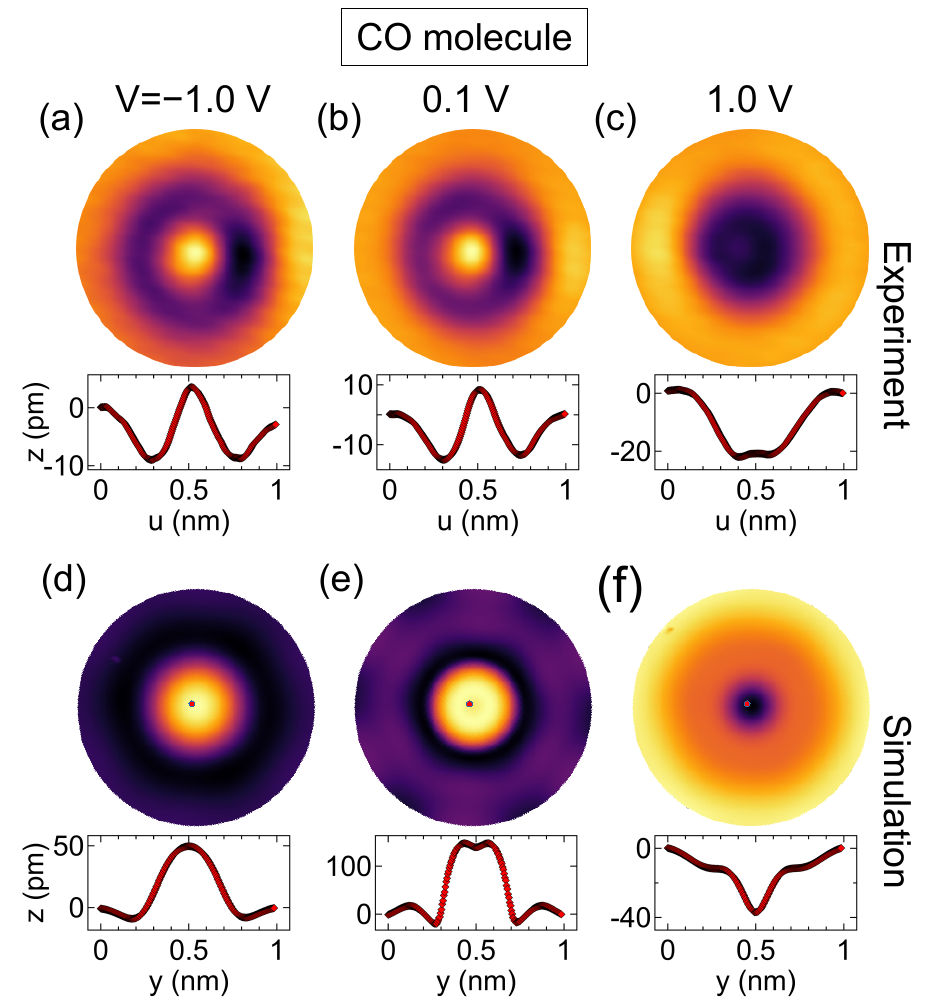}
\caption{(a)--(c) Constant-current STM images of an adsorbed CO molecule on Cu(111) recorded with a straight CO tip at the indicated sample voltages ($50\,\text{pA}$)\@.
The color scales cover apparent heights between $0$ (black) and (a) $18\,\text{pm}$ and (b),(c) $31\,\text{pm}$ (yellow)\@.
(d)--(f) Simulated constant-current STM images of the experimental junction considered in (a)--(c)\@.
An $s$-wave tip is used in (d) and (f) with initial tip--adsorbate distance of (d) $z=700\,\text{pm}$ and (f) $z=800\,\text{pm}$, while the DFT-derived CO tip is used in (e) with $z=300\,\text{pm}$.
The color scales cover apparent heights between $0$ (black) and (d) $59\,\text{pm}$, (e) $170\,\text{pm}$, and (f) $42\,\text{pm}$ (yellow)\@.
Dots mark the center of adsorbed CO\@.
The diameter of the STM images is $1\,\text{nm}$ in (a)--(f)\@.
All cross-sectional profiles depict data along a horizontal line through the center of the STM images.}
\label{fig6}
\end{figure}

In an effort to extend the complexity of the junction, an adsorbed CO molecule (Fig.\,\ref{fig1}) was imaged.
Figure \ref{fig6} shows the voltage-dependent results obtained with a straight CO tip in the experiments [Fig.\,\ref{fig6}(a)--(c)] and simulations [Fig.\,\ref{fig6}(d)--(f)]\@.
The adsorbed CO molecule gives rise to a circular protrusion at $-1.0\,\text{V}$ [Fig.\,\ref{fig6}(a)] and at $0.1\,\text{V}$ [Fig.\,\ref{fig6}(b)] that is centered in a wider and likewise circular depression.
The apparent height of the protrusion changes from $\approx 3\,\text{pm}$ at $-1.0\,\text{V}$ to $\approx 12\,\text{pm}$ at $0.1\,\text{V}$, while the depression is $\approx 17\,\text{pm}$ below the apparent height of the Cu(111) surface.
At $1.0\,\text{V}$ [Fig.\,\ref{fig6}(c)], the central protrusion almost disappears.
Its apparent height is $\approx 15\,\text{pm}$ below the apparent Cu(111) surface height.
The diameter of the circular depression stays nearly constant ($\approx 0.60\,\text{nm}$) for the range of applied sample voltages.

\begin{figure}
\centering
\includegraphics[width=0.95\columnwidth]{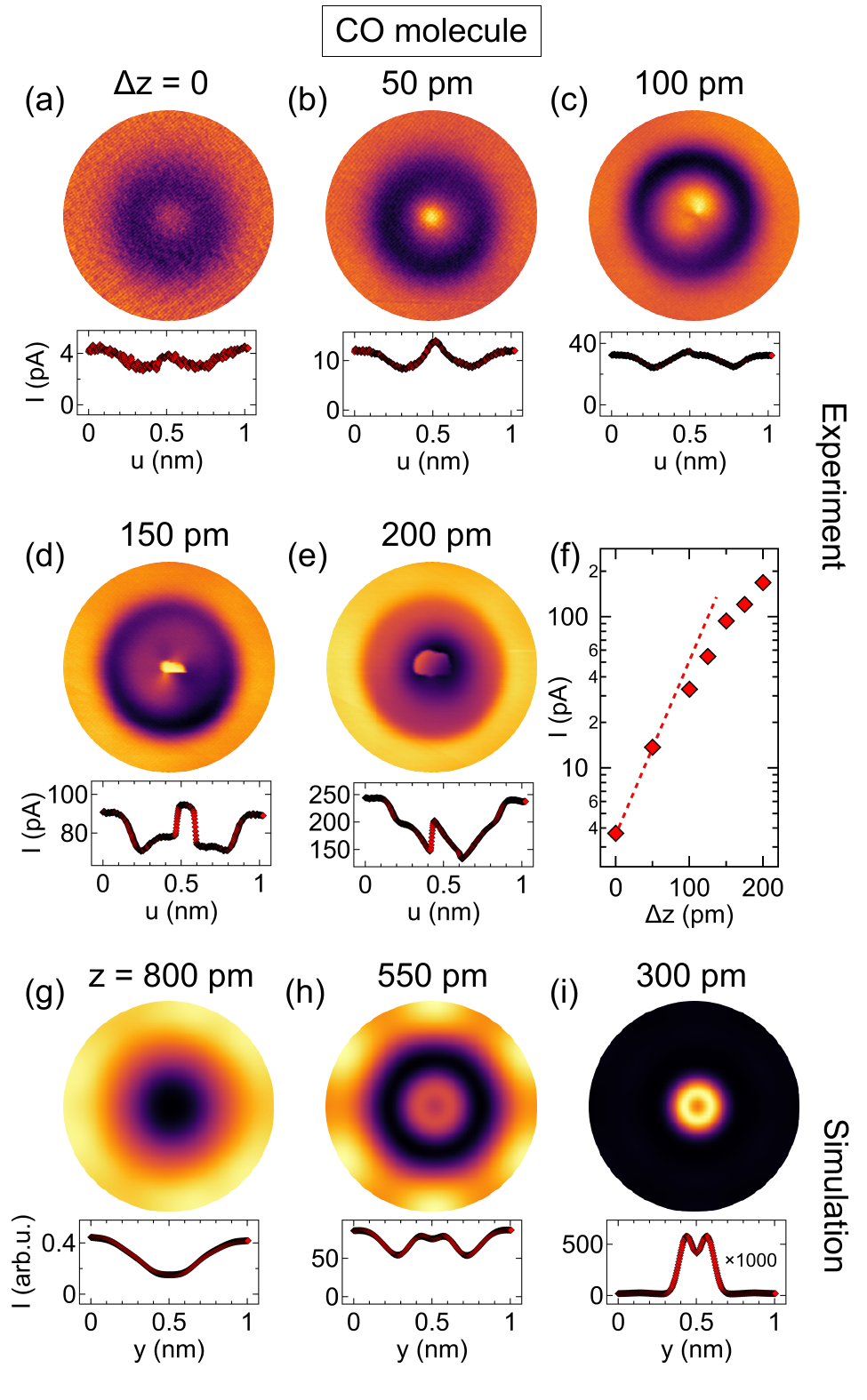}
\caption{(a)--(e) Constant-height current maps of a CO molecule adsorbed on Cu(111) recorded with a straight CO tip at the indicated tip displacements $\Delta z$ ($10\,\text{mV}$)\@.
Zero displacement is defined by the tip--surface distance prior to disabling the feedback loop above Cu(111) at $100\,\text{mV}$ and $50\,\text{pA}$\@.
The color scales (dark to bright) cover currents from (a) $2.3\,\text{pA}$ to $5.5\,\text{pA}$, (b) $7.8\,\text{pA}$ to $14.2\,\text{pA}$, (c) $23.1\,\text{pA}$ to $36.6\,\text{pA}$, (d) $65.9\,\text{pA}$ to $95.5\,\text{pA}$, (e) $124.3\,\text{pA}$ to $253.0\,\text{pA}$\@.
(f) Current $I$ as a function of $\Delta z$ with the dashed line extrapolating the exponential increase from $0$ to $50\,\text{pm}$ to larger $\Delta z$.
The current represents the average value extracted from a circular area (radius: $10\,\text{pm}$) around the central maxima in the constant-height current maps (a)--(e)\@.
(g)--(i) Simulated constant-height STM images at the indicated distances ($100\,\text{mV}$)\@.
The diameter of the STM images is $1\,\text{nm}$ in (a)--(e) and (g)--(j)\@.
All cross-sectional profiles depict data along a horizontal line through the center of the STM images.}
\label{fig7}
\end{figure}

The experimental data at $\pm 1.0\,\text{V}$ can be best reproduced in the simulations by using an $s$-wave tip.
In particular, the transition of the pronounced central protrusion at $-1.0\,\text{V}$ [Fig.\,\ref{fig6}(d)] to a depression at $1.0\,\text{V}$ [Fig.\,\ref{fig6}(f)] is well captured only with the $s$-wave tip.
The DFT-derived CO tip reproduces the experimental contrast at $-1.0\,\text{V}$ with a larger FWHM (not shown) but is unable to show the depression at $1.0\,\text{V}$\@.
At $0.1\,\text{V}$ [Fig.\,\ref{fig6}(e)] both the $s$-wave and the DFT-derived CO tip provide simulated STM images that are in accordance with experimental data.
The increase of the apparent height from $\approx 59\,\text{pm}$ at $-1.0\,\text{V}$ to $\approx 170\,\text{pm}$ at $0.1\,\text{V}$ reflects the same trend observed in the experiments.

\begin{figure*}
\centering
\includegraphics[width=0.7\textwidth]{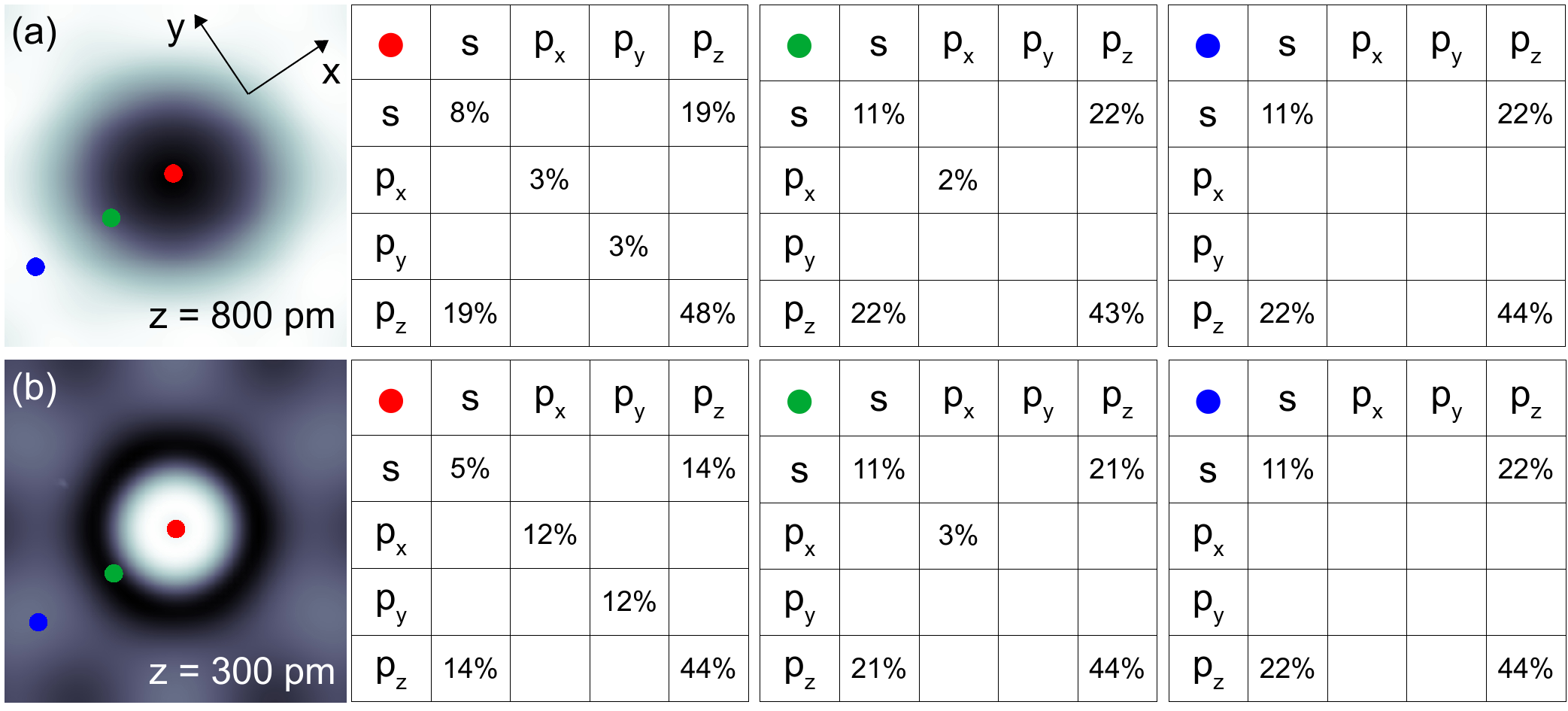}
\caption{Tip orbital decomposition of the tunneling current across a straight CO--CO junction on Cu(111) at $0.1\,\text{V}$ and (a) $z=800\,\text{pm}$, (b) $z=300\,\text{pm}$.
Different lateral tip positions are indicated by the colored dots.
The entries of the $4\times 4$ matrices are fractions of the total tunneling current.
Empty entries reflect vanishing contributions.}
\label{fig8}
\end{figure*}

For adsorbed CO, constant-height mapping of the tunneling current at decreasing tip--surface distances was achieved [Fig.\,\ref{fig7}(a)--(e)], which in the case of the Cu adatom was impeded due to its tip-induced lateral diffusion already at elevated tip--surface separations.
The experimental data show that a decrease of the tip--surface distance from the far tunneling range [$\Delta z=0$, Fig.\,\ref{fig7}(a)] by $\Delta z=50\,\text{pm}$ [Fig.\,\ref{fig7}(b)] leads to a contrast increase of the central protrusion from $3.7\,\text{pA}$ to $13.7\,\text{pA}$\@.
Concomitantly, the contrast due to the Cu(111) surface increases only from $4.4\,\text{pA}$ to $11.9\,\text{pA}$\@.
The FWHM of the central protrusion, however, stays nearly constant ($\approx 0.16\,\text{nm}$)\@.
Further decrease of the tip--surface separation entails strong changes in the appearance of the adsorbed CO molecule.
The central protrusion has considerably widened [FWHM $\approx 0.30\,\text{pm}$ at $\Delta z=100\,\text{pm}$, Fig.\,\ref{fig7}(c)] and exhibits a substructure with two adjacent maxima separated by a thin low-contrast line.
Most likely, the two CO molecules are already so close to each other that the interaction causes their bending \cite{prl_106_046104,prb_96_085415}.
These relaxations of the junction geometry become more pronounced with decreasing CO--CO distance [Fig.\,\ref{fig7}(d),(e)]\@.
The sharp edges confining the central protrusion are indicative of bond formation between the opposing CO molecules in the junction.
The conjectured junction relaxations leave their fingerprint in current-versus-displacement traces, too [Fig.\,\ref{fig7}(f)]\@.
The depicted data were extracted from the central region of the constant-height current maps.
A deviation from a uniform exponential increase of the current occurs in a tip displacement range $50\,\text{pm}\leq\Delta z\leq 100\,\text{pm}$, which matches well the range where image distortions are first observed.

The simulations reported here do not take such junction relaxations and CO--CO interactions into account.
Therefore, calculations of constant-height STM images are restricted to the tunneling range of electrode separations.
In addition, the comparison of simulated and experimental constant-height STM data is qualitative because of different units used for calculated and experimental tunneling currents.
In the far tunneling range ($z=800\,\text{pm}$), the adsorbed CO molecule appears as a depression [Fig.\,\ref{fig7}(g)], which at smaller tip--surface distances [$z=550\,\text{pm}$, Fig.\,\ref{fig7}(h) and $z=300\,\text{pm}$, Fig.\ref{fig7}(i)] is transformed into an increasingly  sharp central protrusion embedded in the wider depression.
This contrast evolution can be well compared with the experimental observations in the tunneling range [$\Delta z=0$ and $\Delta z=50\,\text{pm}$ in Fig.\,\ref{fig7}(a),(b)]\@.
The sharp central indentation of contrast is most likely due to the overlap of $p_x$,$p_y$-orbitals in the CO--CO junction at small distance reflecting the node of the $p$-orbitals.
A lateral displacement of the CO probe from the aligned CO--CO configuration then leads to a finite overlap of these orbitals and the concomitant increase of the tunneling current.
The orbital decomposition of the current (Fig.\,\ref{fig8}, CO position, red dot) indeed shows that the contribution of $p_x$,$p_y$-orbitals becomes indeed larger at small CO--CO separations.

For two tip--surface distances ($z=800\,\text{pm}, 300\,\text{pm}$), the orbital composition of the tunneling current was computed for the straight DFT-derived CO tip (Fig.\,\ref{fig8})\@.
In the far tunneling range of tip--surface separations [Fig.\,\ref{fig8}(a)], the $p_z$-orbital of the O atom gives the dominant direct contribution to the tunneling current at all considered tip positions [dots in Fig.\,\ref{fig8}(a)]\@.
Likewise, $s$/$p_z$ interference terms contribute strongly.
Small amounts of the tunneling current are carried by the $s$, $p_x$, $p_y$-orbitals atop the CO center.
They essentially even vanish upon laterally positioning the CO-terminated tip to the sides of the adsorbed CO molecule.
While $p_x$, $p_y$-orbitals exhibit the highest projected DOS of the O atom at $E_\text{F}$, the orientation of the $p_z$-orbital and its vacuum decay obviously play the key role in determining the magnitude of the tunneling current.
At decreased tip--surface distance [$z=300\,\text{pm}$, Fig.\,\ref{fig8}(b)], the $p_x$, $p_y$-orbitals contribute more strongly to the tunneling current, at the expense of the $s/p_z$ interference terms.

Before concluding it is noteworthy that simulations with a tilted CO tip and upright adsorbed CO were likewise performed (not shown)\@.
However, agreement between experimental and calculated data was difficult to achieve with this assumed junction geometry. 
Therefore, the weak-coupling limit used for the modeling reported here is not applicable to these junctions.
Rather, an increased CO--CO interaction becomes important, which is evidenced by the experimental data.

\section{Conclusion}

The Chen's derivative rules have successfully been applied to describe experimental STM images acquired with a CO-terminated tip of an adsorbed Cu atom and CO molecule on Cu(111)\@.
For the adatom junction, the dependence of topographies on voltage and CO probe tilt angle can be traced to the orientation of specific tip orbitals and their overlap with the wave function of the adatom-surface hybrid.
In case of the CO--CO junction, the impact of the imaging voltage as well as the CO--CO distance are captured in the weak-interaction limit and for the straight geometry.
In all junctions, interference terms resulting from $s$, $p_z$-orbitals are important for describing experimental STM images.
The comparison of measured and simulated STM data with functionalized tips enables the estimation of the tilt angle of the decorated probe, which provides valuable information for the overlap of tip and surface wave functions.  
The latter is important for the interpretation of highly resolved orbital images, the efficiency of inelastic electron tunneling as well as the probing of magnetic exchange interaction.

\begin{acknowledgments}
Financial support from the National Research Development and Innovation Office of Hungary (NKFIH, Grant No.\ FK124100), a Stipendium Hungaricum Scholarship of the Tempus Public Foundation, the János Bolyai Research Grant of the Hungarian Academy of Sciences (Grant No.\ BO/292/21/11), the New National Excellence Program of the Ministry for Culture and Innovation from NKFIH Fund (Grant Nos.\ ÚNKP-22-5-BME-282, ÚNKP-23-5-BME-12) and from the Deutsche Forschungsgemeinschaft through KR 2912/17-1 and KR 2912/21-1 is acknowledged.
\end{acknowledgments}

\appendix*
\section{Exemplary comparison of experimental and simulated data}

\begin{figure}
\centering
\includegraphics[width=0.95\columnwidth]{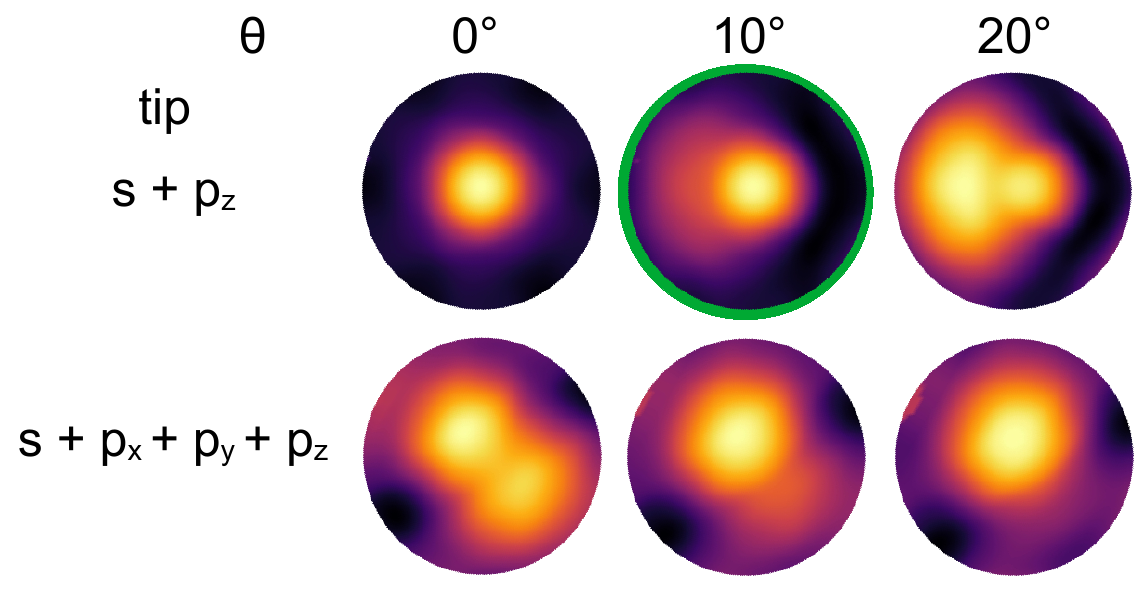}
\caption{Simulated constant-current ($I_0$) STM images acquired with an $(s+p_z)$-wave and $(s+p_x+p_y+p_z)$-wave tip atop a Cu adatom on Cu(111) at the indicated tilt angles.
The tip--surface distance ranges between $610\,\text{pm}$ and $732\,\text{pm}$ ($0^\circ$), $612\,\text{pm}$ and $728\,\text{pm}$ ($10^\circ$), $620\,\text{pm}$ and $721\,\text{pm}$ ($20^\circ$) for the $(s+p_z)$-wave tip and between $671\,\text{pm}$ and $790\,\text{pm}$ ($0^\circ$), $664\,\text{pm}$ and $807\,\text{pm}$ ($10^\circ$), $664\,\text{pm}$ and $818\,\text{pm}$ ($20^\circ$) for the $(s+p_x+p_y+p_z)$-wave tip.
The best match with the experimental data of Fig.\,\ref{fig2}(b) is encircled.}
\label{fig9}
\end{figure}

\begin{figure}
\centering
\includegraphics[width=0.95\columnwidth]{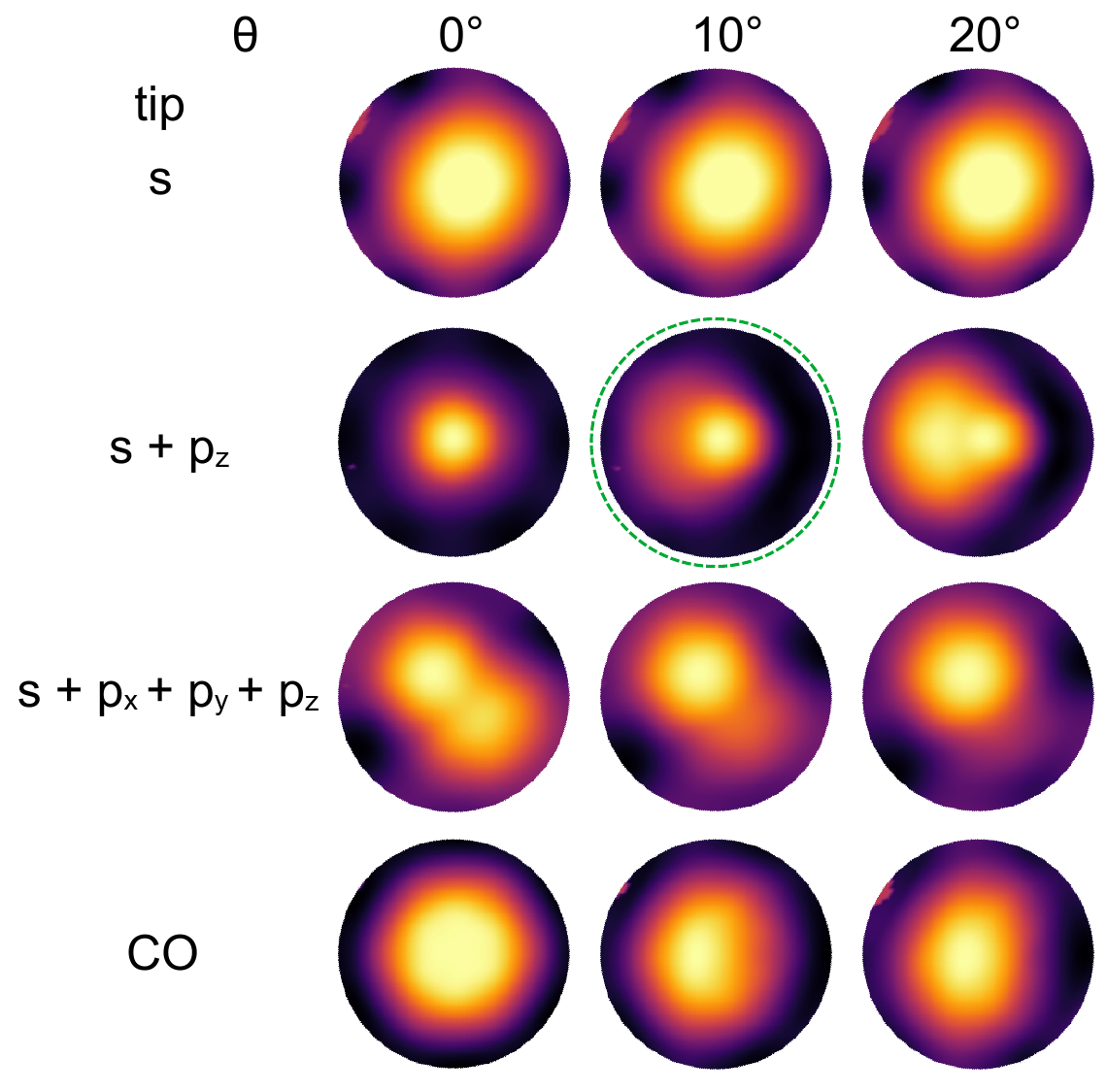}
\caption{As Fig.\,\ref{fig9}, for $10\cdot I_0$ and, additionally, $s$-wave and DFT-derived CO tips.
The rotation of the $s$-wave tip does not change the images, as expected.
The tip--surface distance ranges between $679\,\text{pm}$ and $818\,\text{pm}$ for the $s$-wave tip, between $516\,\text{pm}$ and $654\,\text{pm}$ ($0^\circ$), $517\,\text{pm}$ and $650\,\text{pm}$ ($10^\circ$), $515\,\text{pm}$ and $640\,\text{pm}$ ($20^\circ$) for the $(s+p_z)$-wave tip, between $576\,\text{pm}$ and $704\,\text{pm}$ ($0^\circ$), $574\,\text{pm}$ and $720\,\text{pm}$ ($10^\circ$), $574\,\text{pm}$ and $731\,\text{pm}$ ($20^\circ$) for the $(s+p_x+p_y+p_z)$-wave tip and between $700\,\text{pm}$ and $774\,\text{pm}$ ($0^\circ$), $697\,\text{pm}$ and $794\,\text{pm}$ ($10^\circ$), $691\,\text{pm}$ and $814\,\text{pm}$ ($20^\circ$) for the DFT-derived CO tip. 
The dashed line encircles simulated data that still match the experimental data of Fig.\,\ref{fig2}(b) very well.}
\label{fig10}
\end{figure}

\begin{figure}
\centering
\includegraphics[width=0.95\columnwidth]{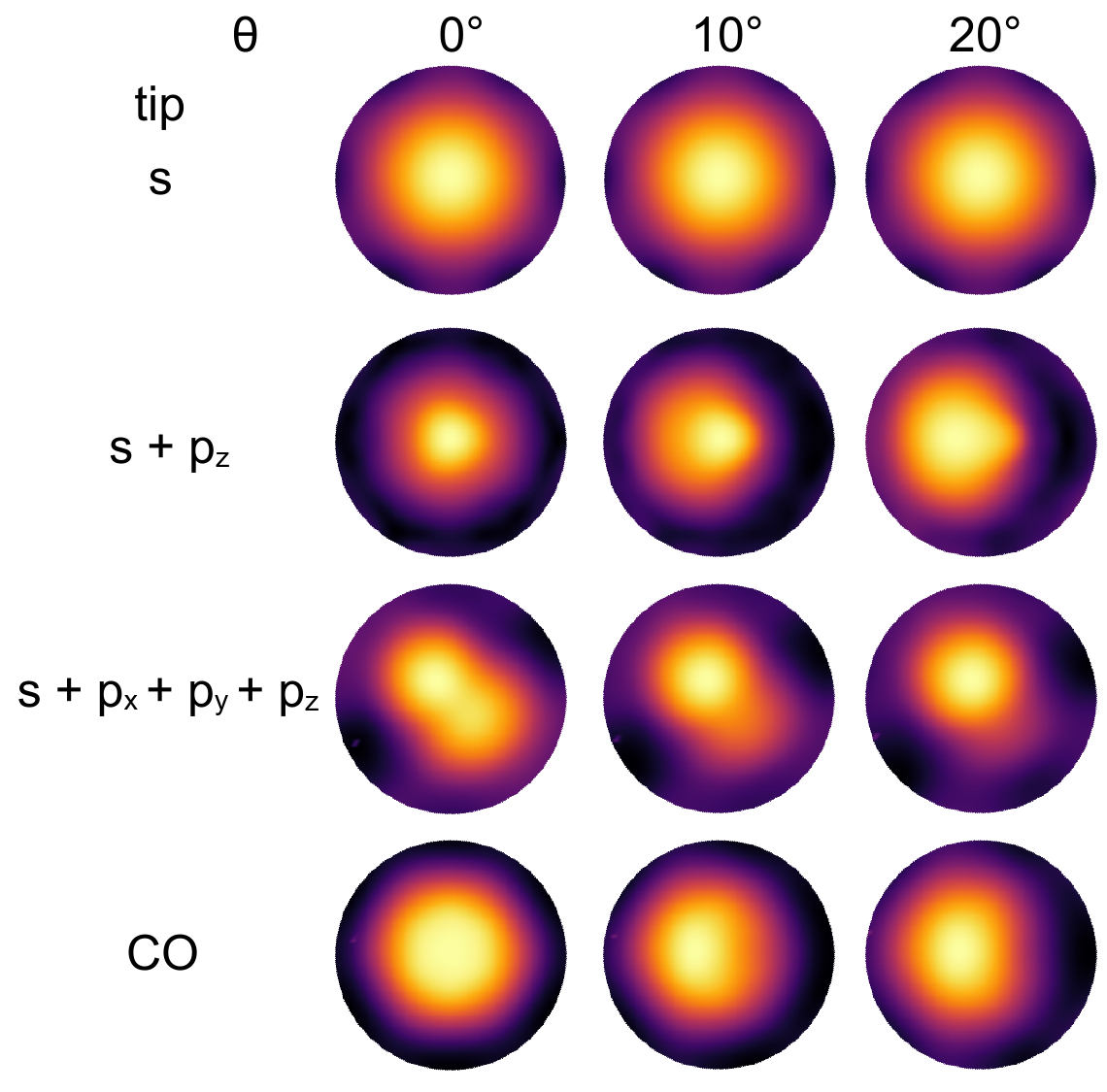}
\caption{As Fig.\,\ref{fig10}, for $100\cdot I_0$\@.
The tip--surface distance ranges between $571\,\text{pm}$ and $736\,\text{pm}$ for the $s$-wave tip, between $412\,\text{pm}$ and $562\,\text{pm}$ ($0^\circ$), $415\,\text{pm}$ and $559\,\text{pm}$ ($10^\circ$), $415\,\text{pm}$ and $562\,\text{pm}$ ($20^\circ$) for the $(s+p_z)$-wave tip, between $486\,\text{pm}$ and $622\,\text{pm}$ ($0^\circ$), $485\,\text{pm}$ and $636\,\text{pm}$ ($10^\circ$), $486\,\text{pm}$ and $646\,\text{pm}$ ($20^\circ$) for the $(s+p_x+p_y+p_z)$-wave tip and between $598\,\text{pm}$ and $690\,\text{pm}$ ($0^\circ$), $597\,\text{pm}$ and $706\,\text{pm}$ ($10^\circ$), $593\,\text{pm}$ and $724\,\text{pm}$ ($20^\circ$) for the DFT-derived CO tip.}
\label{fig11}
\end{figure}

\begin{figure}
\centering
\includegraphics[width=0.95\columnwidth]{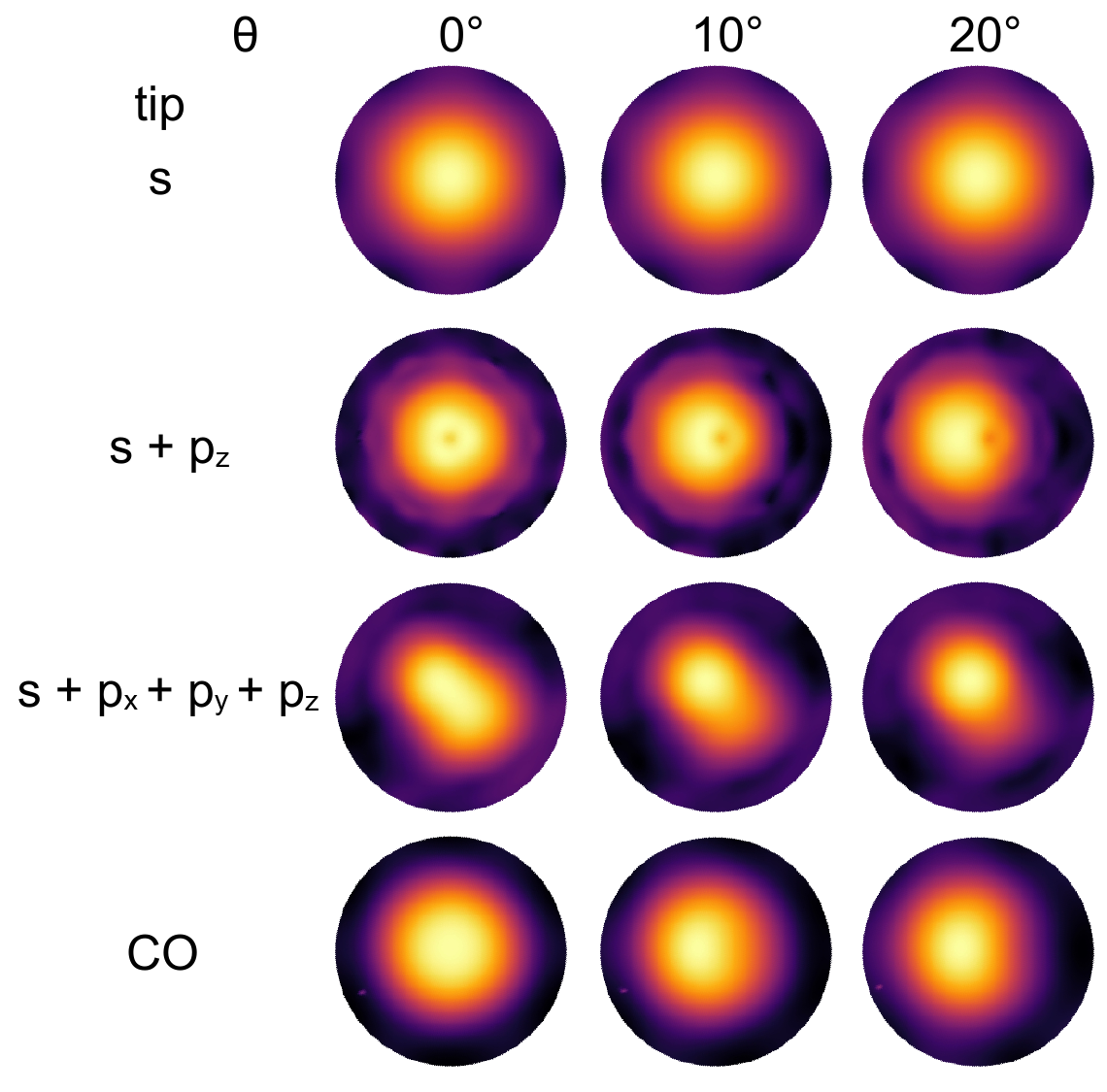}
\caption{As Fig.\,\ref{fig11}, for $1000\cdot I_0$\@.
The tip--surface distance ranges between $470\,\text{pm}$ and $644\,\text{pm}$ for the $s$-wave tip, between $298\,\text{pm}$ and $471\,\text{pm}$ ($0^\circ$), $308\,\text{pm}$ and $479\,\text{pm}$ ($10^\circ$), $319\,\text{pm}$ and $494\,\text{pm}$ ($20^\circ$) for the $(s+p_z)$-wave tip, between $394\,\text{pm}$ and $541\,\text{pm}$ ($0^\circ$), $394\,\text{pm}$ and $551\,\text{pm}$ ($10^\circ$), $394\,\text{pm}$ and $560\,\text{pm}$ ($20^\circ$) for the $(s+p_x+p_y+p_z)$-wave tip and between $500\,\text{pm}$ and $610\,\text{pm}$ ($0^\circ$), $500\,\text{pm}$ and $623\,\text{pm}$ ($10^\circ$), $498\,\text{pm}$ and $638\,\text{pm}$ ($20^\circ$) for the DFT-derived CO tip.}
\label{fig12}
\end{figure}

The galleries of simulated constant-current STM images shown here illustrate the identification of the best match between experimental and calculated data.
To this end, the STM image of Fig.\,\ref{fig2}(b), i.\,e., the constant-current data of a Cu adatom on Cu(111) acquired with a tilted CO tip,  is chosen as an example.
Calculations for a current $I_0$ (Fig.\,\ref{fig9}), $10\cdot I_0$ (Fig.\,\ref{fig10}), $100\cdot I_0$ (Fig.\,\ref{fig11}), and $1000\cdot I_0$ (Fig.\,\ref{fig12}) are presented for different CO tip models ($s$, $s+p_z$, $s+p_x+p_y+p_z$, DFT-derived CO tip) and rotation angles $\theta$.
For the $I_0$ data (Fig.\,\ref{fig9}), the $s$-wave and DFT-derived CO tips have been omitted because the given current value is outside the region of the calculations in terms of tip--surface distance.
In all images, the $y$ direction coincides with a compact $\langle 11\bar{2}\rangle$ direction of the Cu(111) lattice and is horizontal (Fig.\,\ref{fig1})\@.

In Fig.\,\ref{fig9} the best match with the experimental data of Fig.\,\ref{fig2}(b) is encircled.
It is obtained for the $(s+p_z)$-wave tip at a rotation angle of $10^\circ$\@.
The same tip gives rise to a good match with the experimental data at the elevated current $10\cdot I_0$ (dashed circle in Fig.\,\ref{fig10}), while for more elevated currents (Figs.\,\ref{fig11},\ref{fig12}) deviations occur.

The simulations using a $(s+p_x+p_y+p_z)$-wave tip give rise to two protrusions aligned along a direction that encloses $45^\circ$ with the horizontal.
These protrusions are most pronounced for $\theta=0^\circ$ and weaken with increasing tilt angle and current.
Most likely, these features result from the overlap of the $p_x$ and $p_y$ tip orbitals with in-plane orbitals of the top substrate atomic layer.

%\bibliography{ref}

\end{document}